\documentclass{ifacconf}

\usepackage{graphicx}      
\usepackage{natbib}        
\usepackage{epsfig} 
\usepackage{amsmath}
\usepackage{amssymb} 
\usepackage{subcaption}
\usepackage{mhchem}
\usepackage{array}
\usepackage{multirow}
\usepackage{xcolor}
\begin{document}
\begin{frontmatter}

\title{A Modified Rao-Blackwellised Particle Filter Based on Updated Process Noise Covariance for Chemical Reaction Networks}


\author[First]{Suryasnata Dash} 
\author[First]{Abhishek Dey} 

\address[First]{National Institute of Technology, 
   Rourkela, Odisha, India, 769008 (e-mail: \{523ee1001, deyab\}@nitrkl.ac.in).}

\begin{abstract}                
It is necessary to characterize chemical reaction networks (CRNs) accurately to facilitate engineering of both synthetic and naturally occurring biological systems. Estimation of multiple unknown parameters and states in CRNs requires first, an identifiability analysis, and then adoption of a suitable estimation technique. In this paper, we took an example of a reduced order gene expression system, performed parameter sensitivity analysis for multiple unknown parameters, and used a modified Rao-Blackwellised particle filter (RBPF) to estimate parameters and states. The proposed framework estimates parameters using a particle filter and states using an extended Kalman filter (EKF) with process noise covariance updated recursively based on chemical Langevin equation (CLE). We demonstrate finite filter deviation of proposed RBPF from the conventional RBPF involving multiple Kalman filters theoretically. We compared the accuracy of parameter estimation, error in state estimation, and whiteness of the innovation sequence for the proposed filter with conventional RBPF, previously implemented continuous-discrete EKF in addition to fixed choices of noise covariance.  We found that the RBPF with updated noise covariance finds a balance between these three criteria as compared to other applied filters, demonstrating its suitability for joint state and parameter estimation for stochastic CRNs.
\end{abstract}

\begin{keyword}
State and parameter estimation, Identifiability analysis, Particle filtering, Nonlinear systems, Biological systems.
\end{keyword}

\end{frontmatter}

\section{INTRODUCTION}
Chemical reaction networks (CRN) represent biomolecular systems using chemical reactions and conservation laws. Mathematical modeling of CRNs is required for engineering new biological systems or analyzing existing ones. CRNs are mathematically represented through ordinary differential equations (ODE) based on mass conservation law. Chemical master equation (CME) gives a probabilistic structure of the system equations but faces computational issues due to high dimensions (\cite{gillespie1977exact}). As an alternative, chemical Langevin equation (CLE) can be used for stochastic differential equation (SDE)-based representation of CRNs (\cite{gillespie2000chemical}). Often there are multiple unknown parameters in these mathematical models, and Bayesian estimation frameworks are typically used to estimate them. Since noise affects biomolecular systems intrinsically and extrinsically, along with its nonlinear dynamics, frameworks like extended Kalman filter (EKF) (\cite{sun2008extended, meskin2011parameter, Lillacci2010}) and unscented Kalman filters (UKF) (\cite{baker2011unscented, meskin2013parameter}) are used. While these methods are computationally easier, Kalman filtering handles Gaussian noise. For non-Gaussian noise and multi-parameter estimation, particle filters (PF) can be used for state and parameter estimation. A comparison of EKF, UKF and PF was done for estimating state and parameters of temperature based cellular response in \textit{E. coli} in~\cite{liu2012state}. The states and parameters of heart rate model was estimated using PF (\cite{du2019model}). Parameter estimation in gene regulatory networks is done using PF with resampling based on Metropolis-Hastings acceptance criteria in~\cite{shen2010inferring}. Joint state and parameter estimation have been implemented through PFs shown in~\cite{Rathinam2021} and have also been applied in reduced approximations of stochastic reaction networks in~\cite{fang2020stochastic}. Block PF have also been used for state estimation in reaction-diffusion systems (\cite{magalhaes2023block}). However, a recurring problem in estimation with conventional PFs is particle degeneracy. To overcome this problem, a  variant of PF, regularized particle filter (RPF) is used in efficient estimation of states in biomolecular networks in~\cite{fang2023convergence}. Another variant, Rao-Blackwellised particle filter (RBPF), where a subset of state is marginalized out for closed form solution to reduce high dimension problem and improve performance can also be used. Early application of RBPF considered $N$ Kalman filters for $N$ particles (\cite{li2004estimation}). This increases computational burden. The algorithm was improved where only parameters were considered for particle filtering and Kalman filters were implemented through mean estimates (\cite{mustiere2006modified}).  \cite{lindsten2012non} also estimated nonlinear states and static parameters conditional on nonlinear state distribution using Rao-blackwellised marginal particle filter (RBMPF). System parameters are estimated using Rao-Blackwellisation approach where Zakai equations are used for parameter marginalization and estimation (\cite{fang2024effective}). 

In this paper, we consider a modified RBPF framework to estimate multiple unknown parameters in a CRN. Within this framework, the parameters are estimated using a particle filter and states are estimated using EKF similar to approaches in~\cite{li2004estimation} and~\cite{mustiere2006modified} but with updated process noise covariance based on CLE. Previously, an updating process noise covariance was implemented for EKF using CLE to estimate states and parameters (\cite{dey2019kalman}). This approach was also extended to reduced order models of biomolecular systems (\cite{dash2025extended}). The contributions of the paper are,
\begin{itemize}
	\item We propose a CLE-based, estimate-dependent process noise ocvariance into the modified RBPF, updating it at each time step using posterior state and parameter estimates to account for the state- and parameter-dependent stochasticity of CRNs. 
	\item We provide theoretical outline on inter-algorithm discrepancy between conventional and modified RBPF in terms state propagation, error covariance propagation, likelihood, weight and estimated parameters.
\end{itemize}
  We provide a comparison of estimation and filter performance between conventional RBPF (\cite{li2004estimation}), continuous-discrete EKF (\cite{dash2025extended}) and our own implemented RBPF with estimate-dependent process noise covariance. We found that the proposed RBPF with updated process noise covariance achieves a balance between the accuracy of particle filter for parameter estimation and optimality of the EKF for state estimation while compared to fixed choices of process noise covariance. Comparison with conventional filter and CD-EKF (\cite{dash2025extended}) show that while some aspects are comparable or better in case modified RBPF and other aspects in conventional RBPF, CD-EKF, overall analysis shows modified RBPF with estimate-dependent process noise covariance to be a better choice for state and parameter estimation.


\section{SYSTEM DESCRIPTION AND APPLIED FILTERS}
This section elaborates on the system model, established filters used for estimation of states and parameters, eventually stating the difference between our modified RBPF from other established filters.
\subsection{System model}A discrete-time nonlinear state-space model is given as,
\begin{align*}
	\zeta_{k+1}&= f(\zeta_{k}, \theta_k) + G_k\mathbf{W}_{k}, \\ 
	y_{k+1} &= C\zeta_{k+1} + \mathbf{v}_{k+1}, \tag{1} \label{sys_eqn}
\end{align*}
where $\zeta_k \in \mathbb{R}^{n\times 1}$ and $\theta_k \in \mathbb{R}^{m\times 1}$ are state and parameter vectors respectively for $k^{th}$ time step whereas $y_{k+1} \in \mathbb{R}^{p\times 1}$. $G_k$ represents the noise coefficient matrix for state. \(\mathbf{W}_{k} \sim \mathcal{N}(0, Q_0)\) and \(\mathbf{v}_{k+1} \sim \mathcal{N}(0, R)\) are independent Gaussian process and measurement noise processes where $Q_0 = I_{n \times n}$. 
\subsection{Estimate-dependent process noise covariance} We implement an estimate-dependent updating process noise covariance in EKF part of the modified RBPF. The system equation with sampling period $\Delta$ based on CLE is,
\[ \zeta_{k+1} = \zeta_k + \Delta(V\mathcal{A}(\zeta_k,\theta_k)) + V diag(\sqrt{\Delta(\mathcal{A}(\zeta_k,\theta_k))})\mathbf{W}_k, \tag{11} \label{cle_eqn}\]
where $V$ is stoichiometric matrix and $\mathcal{A}(\zeta_k,\theta_k)$ is the propensity vector (\cite{gillespie2000chemical}). Using~\eqref{cle_eqn} for state prediction, process noise covariance matrix is updated using $G_k = V diag(\sqrt{\Delta(\mathcal{A}(\zeta^+_{k},\theta^+_{k}))})$ which has previous state and parameter estimates dependency. The updated process noise covariance is represented as,
\[ \hat{Q}_k = G_kQ_0G_k^T \implies \hat{Q}_k = \Delta V diag(\mathcal{A}(\zeta^+_{k},\theta^+_{k}))V^T. \tag{12} \label{updated_Q_eqn}\]
\subsection{EKF} EKF is used to estimate states for nonlinear systems. It follows prediction for updating the prior information,
\begin{gather*}
	\zeta^-_{k+1} = f(\zeta^+_k,\mathbf{\theta}_k) \\
	P^-_{k+1} = A_k P^+_k (A_k)^\top + Q_f, \tag{2} \label{ekf_priori_eqn}
\end{gather*}
and estimation step for updating the posterior information based on recent measurement,
\begin{gather*}
	\zeta^+_{k+1} = \zeta^-_{k+1} + K_{k+1}(y_{k+1} - C\zeta^-_{k+1}) \\
	P^+_{k+1} = (I - K_{k+1} C)P^-_{k+1}(I - K_{k+1} C)^T + K_{k+1}RK^T_{k+1} \\	
	K_{k+1} = P^-_{k+1}C^TS_{k+1}^{-1}, \tag{3} \label{ekf_posterior_eqn}
\end{gather*}
where $S_{k+1} = CP^-_{k+1}C^T + R$ and $Q_f$ denotes the process noise covariance used in EKF, which can be a constant matrix or updated based on estimates ($\hat{Q}_k$).
\subsection{Conventional RBPF} RBPF is used for state and parameter estimation based on measured data, combining Kalman filter and PF methodologies to decrease the dimensionality of PF while maintaining performance. The conventional RBPF(\cite{li2004estimation}) implements $N$ EKFs where $N$ denotes particle size. The state and parameter particles are estimated based on posterior probability, 
\[ p(\zeta_k,\theta_k \mid y_{1:k}) =  p(\zeta_k \mid \theta_k, y_{1:k})p(\theta_k \mid y_{1:k}), \]
where conditional posterior \(p(\zeta_k \mid \theta_k, \mathbf{y}_{1:k})\) is marginalized and implemented through $N$ EKFs using~\eqref{ekf_posterior_eqn} simultaneously. Particle filtering is used for $\theta_k$ calculation. The $i^{th}$ particle weight is updated using likelihood function:
\[
w_k^{i} \propto p(\mathbf{y}_k \mid \theta_k^i, \zeta_{k|k-1}^{i}), \tag{4} \label{pf_weight_eqn}
\]
where $\zeta_{k|k-1} = \zeta^-_{k}$ is updated by EKF and gets normalized to calculate posterior parameter,
\[ \theta_k = \sum_{i=1}^{N}\hat{w}_k^i
\theta_k^i. \tag{5} \label{pf_posterior_param_eqn}\] 
The time update of resampled parameter particles utilize posterior parameter values, where $i^{th}$ parameter particle is sampled from probability distribution using a kernel smoothing approach,
\[ p(\theta_{k+1}^i \mid \theta_{k}^i) = \mathcal{N}(\theta_{k}^i, \Sigma_k). \tag{6} \label{pf_prior_param_eqn} \] 
Here, $\Sigma_k$ is the parameter covariance which is a positive definite matrix with small diagonal values and $\theta_{k}$ is the posterior parameter value of previous time instant. 
\subsection{Modified RBPF} We modify the RBPF similar to~\cite{mustiere2006modified} with a single EKF based on updating process noise covariance using weighted mean posterior states and parameters to update. State propagation is updated based on parameter particles as the propagation is affected for different particle values. Initial parameter particles are uniformly distributed and corresponding states to be affected have same initial values. For prediction step a prior state $\zeta^{i-}_{k+1}$ is updated based on given prior parameter particle $\theta^{i-}_{k+1}$ and equation can be updated as,
\[ \zeta^{i-}_{k+1} = f(\zeta^{i+}_k, \theta^{i-}_{k+1}). \tag{7} \label{mrbpf_priori_state_eqn} \]
Priori error covariance is updated based on weighted mean of posterior state and parameter vector. The priori error covariance is calculated with estimate-dependent process noise covariance~\eqref{updated_Q_eqn} as, 
\begin{gather*}
	P^-_{k+1} = A_kP^+_{k}A^T_k + \hat{Q}_k, \ \text{where} \  A_k = \frac{\partial f(\zeta^+_{k}, \theta^+_{k})}{\partial\zeta^+_{k}} \tag{8} \label{mrbpf_priori_cov_eqn}
\end{gather*}
Weights are updated based on~\eqref{pf_weight_eqn}, and normalized to calculate posterior parameter $\theta^+_{k+1}$ given in~\eqref{pf_posterior_param_eqn}. Stratified resampling is applied using adaptive resampling with effective sample size, $N_{\text{eff}} = 1/\sum_{i=1}^N \left(\hat{w}^i \right)^2$(\cite{martino2017effective}) to obtain $\theta^{i+}_{k+1}$ and posterior state information is updated as,
\[ \zeta^{i+}_{k+1} = \zeta^{i-}_{k+1} + K_{k+1}(y_{k+1} - C\zeta^{i-}_{k+1}) \tag{9} \label{mrbpf_prosterior_state_eqn}\] 
while Kalman gain and posterior error covariance is same as mentioned in ~\eqref{ekf_posterior_eqn}. The next prior particles are updated using random walk propagation technique,
\[ \theta^{i-}_{k+2} \sim \mathcal{N}(\theta^{i+}_{k+1}, \Sigma_{k+1}). \tag{10} \label{mrbpf_param_priori_eqn}\]
The particle covariance is set as $\Sigma_{k+1} = 10^{-((k+1)-1)}{\Sigma}^+_{k+1}$ where ${\Sigma}^+_{k+1}$ denotes the posterior particle covariance. There is reduction of covariance with each iteration and eventually gets maintained by a threshold covariance $\Sigma_{lb}$. As biological parameters may also be small in range, using kernel smoothing with parameter shrinkage~\cite{liu2001combined} reduces particle exploration and may be unable to reach near the true state due to continuous covariance reduction. While covariance reduction is necessary to prevent overdispersion of parameter particles, it is also necessary to maintain a threshold variance which would allow to explore in case of significantly uncertain priori. 


\section{QUANTITATIVE DIFFERENCE BETWEEN CONVENTIONAL AND MODIFIED RBPF}
As our approach to RBPF implements a single EKF in contrast to the conventional RBPF, we compare posterior parameter discrepancy between two algorithms where both algorithms utilize updating process noise covariance. This is also an attempt to give an approximation of modified RBPF to conventional one, in terms of estimation performance. The theoretical comparison is found based on following assumptions.
\begin{itemize}
	\item[1.] A nonlinear function $f(x,\theta)$ is Lipschitz continuous and differentiable satisfying $\|f(x_1,\theta_1) - f(x_2,\theta_2)\| \le L_f\|x_1 - x_2\| + L_{\theta f}\|\theta_1 - \theta_2\|$, along with its Jacobian $A = \frac{\partial f(x,\theta)}{\partial x}$ being Lipschitz continuous satisfying $\|A(x_1,\theta_1) - A(x_2,\theta_2)\| \le L_A\|x_1 - x_2\| + L_{\theta A}\|\theta_1 - \theta_2\|$.
	\item [2.] The propensity function $\mathcal{A}(x, \theta)$ is Lipschitz continuous satisfying, $\|\mathcal{A}(x_1, \theta_1) - \mathcal{A}(x_2, \theta_2)\| \le L_{\mathcal{A}}\|x_1 - x_2\| + L_{\theta\mathcal{A}}\|\theta_1 - \theta_2\|$.
	\item[3.] Parameter particles for both conventional and modified RBPF and their corresponding posterior parameters belong to compact sets $S_{\theta_c}$ and $S_{\theta_M}$ respectively.
	\item[4.] Initial parameter particles, state and posterior error covariance are same for both algorithms.
\end{itemize}
Both algorithms follow same state and particle propagation formulae, and as per Assumption$4$, initial state and parameter particles remain same, then with particle covariance update, subsequent covariance will reduce in size, from which subsequent particles will be sampled. Then, the particle sampling will be within the initial covariance range and even if both algorithms will have different subsequent particles, they will be bounded by $\|\theta^i_{c} - \theta^i_{M}\| \le \sigma_0$, where $\theta^i_{c}$ represents parameter particle in conventional RBPF, $\theta^i_{M}$ is a parameter particle in modified RBPF, $\sigma_0 > 0$ and finite, representing initial deviation of particles. The subscript $(.)_c$ represents the variable(s) in conventional RBPF, $(.)_M$ represents variable(s) for modified RBPF in the following analyses.
\subsection{Posterior parameter discrepancy}
\begin{lem}\label{lem:priori_eqn}
	The priori state and error covariance defined in conventional and modified RBPF for single step have finite discrepancy.
\end{lem}
\textit{Proof: $(i)$} The priori state depends on posterior state and both algorithms use particle based propagation. Since $\|\zeta^{i-}_{c_k} - \zeta^{i-}_{M_k}\| = \|f(\zeta^{i+}_{c_{k-1}}, \theta^i_{c_k}) - f(\zeta^{i+}_{M_{k-1}}, \theta^i_{M_k})\|$, if we consider $\|\zeta^{i+}_{c_{k-1}} - \zeta^{i+}_{M_{k-1}}\|$ is finite then based on Assumption$1$, error is bounded as,
\[ \|\zeta^{i-}_{c_k} - \zeta^{i-}_{M_k}\| \le L_f\|\zeta^{i+}_{c_{k-1}} - \zeta^{i+}_{M_{k-1}}\| + L_{\theta f}\|\theta^i_{c_k} -  \theta^i_{M_k}\|. \tag{13} \label{err_bound_priori_state} \]
\textit{$(ii)$} The priori error covariance discrepancy $P^{i-}_{c_k} - P^-_{M_k}$ deals with two differences: $A^i_{c_{k-1}}P^{+i}_{c_{k-1}}{A^i}^T_{c_{k-1}} - A_{M_{k-1}}P^+_{M_{k-1}}$ $A^T_{M_{k-1}}$ and $\hat{Q}^i_{c_{k-1}} - \hat{Q}_{M_{k-1}}$. First part involves posterior error covariance and we consider difference between posterior error covariance of both algorithms is finite, $P^{+i}_{c_{k-1}} - P^+_{M_{k-1}} = \epsilon_{P^+}$. First difference can be rewritten as $A^i_{c_{k-1}}(P^+_{M_{k-1}} + \epsilon_{P^+}){A^i_{c_{k-1}}}^T - A_{M_{k-1}}P^+_{M_{k-1}}A^T_{M_{k-1}} + A^i_{c_{k-1}}P^+_{M_{k-1}}A^T_{M_{k-1}} -  A^i_{c_{k-1}}P^+_{M_{k-1}}A^T_{M_{k-1}}$. Rearranging the terms, we get,
\begin{gather*}
	A^i_{c_{k-1}}P^{+i}_{c_{k-1}}{A^i}^T_{c_{k-1}} - A_{M_{k-1}}P^+_{M_{k-1}}A^T_{M_{k-1}} = A^i_{c_{k-1}}\epsilon_{P^+}{A^i}^T_{c_{k-1}} \\ + A^i_{c_{k-1}}P^+_{M_{k-1}}({A^i}^T_{c_{k-1}} - A^T_{M_{k-1}}) + \{(A^i_{c_{k-1}} - A_{M_{k-1}} )P^+_{M_{k-1}} \\ A^T_{M_{k-1}}\}. \hspace*{7.5cm}	
\end{gather*} 
Considering Jacobian matrices of both algorithms and posterior error covariance $P^+_{M_{k-1}}$ to be bounded i.e, $\|A^i_{c_{k-1}}\|, \|A_{M_{k-1}}\| \le c_A$, $\|P^+_{M_{k-1}}\| \le p_{ub}$ along with Assumption$1$ for Lipschitz continuity of Jacobian matrix, the first difference term is represented as, 
\begin{gather*}
	\|A^i_{c_{k-1}}P^{+i}_{c_{k-1}}{A^{i^T}_{c_{k-1}}} - A_{M_{k-1}}P^+_{M_{k-1}}A^T_{M_{k-1}}\| \le c^2_A\|\epsilon_{P^+}\| + \\ 2c_Ap(L_A\|\zeta^{i+}_{c_{k-1}} - \zeta^+_{M_{k-1}}\| + L_{\theta A}\|\theta^i_{c_k} -  \theta^i_{M_k}\|). 
\end{gather*} 
The second difference terms involves deviation between estimate-dependent updating process noise covariance of both algorithms, $\hat{Q}^i_{c_{k-1}} - \hat{Q}_{M_{k-1}}$ which can be represented based on CLE form, $\Delta V(diag(\mathcal{A}^i(\zeta^{i+}_{c_{k-1}})) - diag(\mathcal{A}(\zeta^+_{M_{k-1}})))V^T = \Delta V(diag(\mathcal{A}^i(\zeta^{i+}_{c_{k-1}}) - \mathcal{A}(\zeta^+_{M_{k-1}}))V^T$. Based on Assumption$3$, where propensity function is Lipschitz continuous, the error norm is given by \begin{gather*}
\|\hat{Q}^i_{c_{k-1}} - \hat{Q}_{M_{k-1}}\| \le \Delta v^2(L_{\mathcal{A}}\|\zeta^{i+}_{c_{k-1}} - \zeta^+_{M_{k-1}}\| + L_{\theta\mathcal{A}}\|\theta^i_{c_k} \\ - \theta^i_{M_k}\|). \hspace*{6.5cm}
\end{gather*} 
The priori error covariance deviation is represented as,
\begin{gather*}
	\|P^{i-}_{c_k} - P^-_{M_k}\| \le c^2_A\|\epsilon_{P^+}\| + (2c_ApL_A + \Delta v^2L_{\mathcal{A}})\|\zeta^{i+}_{c_{k-1}} - \\ 
  \hspace*{1cm}	\zeta^+_{M_{k-1}}\| + (2c_ApL_{\theta A} + \Delta v^2L_{\theta\mathcal{A}})\|\theta^i_{c_k} -  \theta^i_{M_k}\|.  \tag{14} \label{err_bound_priori_cov_eqn}
\end{gather*}
\begin{lem}\label{lem:log-likelihood}
	The inter-algorithm one step discrepancy of log-likelihood corresponding to each particle is finite.
\end{lem}
\textit{Proof:} The log-likelihood of each state dependent on parameter particle has the representation $l^i = -log(2\pi det(S)^{\frac{p}{2}})$ $+ (y - C\zeta^{i-})^TS^{-1}(y - C\zeta^{i-})$. For both algorithms, log-likelihood functions are,
\begin{gather*}
	l^i_{c_k} = -\frac{p}{2}log(2\pi) -\frac{1}{2}log(det(S^i_{c_k})) - e^i_{c_k} \\
	l^i_{M_k} = -\frac{p}{2}log(2\pi) -\frac{1}{2}log(det(S_{M_k})) - e_{M_k},
\end{gather*}
where $e^i_{c_k} = (y_k - C\zeta^{i-}_{c_k})^T(S^i_{c_k})^{-1}(y_k - C\zeta^{i-}_{c_k})$ and $e_{M_k} = (y_k - C\zeta^{i-}_{M_k})^T(S_{M_k})^{-1}(y_k - C\zeta^{i-}_{M_k})$. The deviation between log-likelihoods will have two differences, $-\frac{1}{2}log(det(S^i_{c_k})) + \frac{1}{2}log(det(S_{M_k}))$ and $-e^i_{c_k} + e_{M_k}$. Considering innovation covariance of both algorithms differ by $\epsilon_S$, the first difference can be rewritten as $-\frac{1}{2}log(det(S_{M_k} + \epsilon_S)) + \frac{1}{2}log(det(S_{M_k}))$. Based on log-determinant perturbations, $-log(det(S_{M_k} + \epsilon_S)) \approx -log(det(S_{M_k})) - trace(S^{-1}_{M_k}\epsilon_S)$. Then, $$\frac{1}{2}(log(det(S_{M_k})) - log(det(S^i_{c_k}))) \approx -\frac{1}{2}(trace(S^{-1}_{M_k}\epsilon_S)).$$
We can find the bound of trace through spectral norm,
$$ trace(S^{-1}_{M_k}\epsilon_S) \le p\|S^{-1}_{M_k}\epsilon_S\| \le \frac{p}{r_{lb}}\|\epsilon_S\|. $$
Since $\epsilon_S = CP^{i-}_{c_k}C^T - CP^-_{M_k}C^T$, we can use $\epsilon_S = C(P^{i-}_{c_k} - P^-_{M_k})C^T$ where $P^{i-}_{c_k} - P^-_{M_k}$ is already known to be bounded. The second difference term $e_{M_k}-e^i_{c_k}$ consists inverse innovation matrices which are positive definite and both $(S^i_{c_k})^{-1}, (S_{M_k})^{-1} \le r_{lb}I$. Considering $a = y_k - C\zeta^{i-}_{c_k}$ and $b = y_k - C\zeta^{i-}_{M_k}$, 
\begin{gather*}
	 e_{M_k} - e^i_{c_k} = b^TS^{-1}_{M_k}b - a^T{S^i}^{-1}_{c_k}a 
	 = b^TS^{-1}_{M_k}b - b^T{S^i}^{-1}_{c_k}b + \\ b^T{S^i}^{-1}_{c_k}b - a^T{S^i}^{-1}_{c_k}a
	 = (b - a)^T{S^i}^{-1}_{c_k}(b + a) + b^T(S^{-1}_{M_k} -\\ {S^i}^{-1}_{c_k})b \hspace*{7.5cm}
\end{gather*}
If innovation errors $\|a\| \le C_a$ and $\|b \|\le C_b$ are finite then $ \|a + b\| \le C_a + C_b$ and $S^{-1}_{M_k} - {S^i}^{-1}_{c_k} = S^{-1}_{M_k}(S^i_{c_k} - S_{M_k}){S^i}^{-1}_{c_k}$ is based on matrix inverse property. Then, 
\begin{gather*}
	\|e_{M_k} - e^i_{c_k}\| \le (C_a + C_b)\|b-a\| \|{S^i}^{-1}_{c_k}\| + C_b^2\|S^{-1}_{M_k}\| \|S^i_{c_k} - \\ S_{M_k}\| \|{S^i}^{-1}_{c_k}\| 
 \le \frac{(C_a + C_b)\|C\|}{r_{lb}}\|\zeta^{i-}_{c_k} - \zeta^{i-}_{M_k}\| + \frac{C_b^2}{r^2_{lb}}\|\epsilon_S\| \\
\end{gather*}
Log-likelihood difference is $l^i_c - l^i_M \le - 0.5(trace(S^{-1}_{M_k}\epsilon_S)) + e_{M_k} - e^i_{c_k}$ and,
\[ \|l^i_c - l^i_M\| \le \frac{c^2p}{r_{lb}}\|P^{i-}_{c_k} - P^-_{M_k}\| + \frac{(C_a + C_b)c}{r_{lb}}\|\zeta^{i-}_{c_k} \]
\begin{gather*}
	 - \zeta^{i-}_{M_k}\| + \frac{C_b^2c^2}{r^2_{lb}}\|P^{i-}_{c_k}- P^-_{M_k}\| \le \frac{(C_a + C_b)c}{r_{lb}}\|\zeta^{i-}_{c_k} \\ - \zeta^{i-}_{M_k}\| + (\frac{0.5c^2p}{r_{lb}} + \frac{C_b^2c^2}{r^2_{lb}})\|P^{i-}_{c_k}- P^{-}_{M_k}\|. \hspace*{1cm} \tag{15} \label{err_bound_loglike_eqn}
\end{gather*}
Since log-likelihood is scalar, $\|l^i_c - l^i_M\| \equiv |l^i_c - l^i_M|$ and~\eqref{err_bound_loglike_eqn} is still valid. Following above lemmas, we present the theorem for posterior parameter discrepancy between two algorithms.

\textit{Theorem 1.} Given a conventional and modified RBPF, the one step discrepancy of posterior parameter before resampling is finite with constants $c_{\theta_c}$, $\epsilon_w$ and $c_{w_M}$.
\[ \|\theta^+_{c_k} - \theta^+_{M_k}\| \le \sum_{i=1}^N c_{\theta_c}(e^{\epsilon_w} + 1) + c_{w_M}\|\theta^i_{c_k} - \theta^i_{M_k}\| \]
if discrepancy in priori state, error covariance update and likelihood between given algorithms is finite. 

\textit{Proof:} We first prove that the inter-algorithm discrepancy of normalized weights for each particle is finite. Since weights are related to log-likelihood, we can also find the deviation between normalized weights for independent particle of both algorithms as $\hat{w}^i = w^i/\sum_{i=1}^{N}w^i$ . Since $l^i_c = log(w^i_c)$ and $l^i_M = log(w^i_M)$, then $l^i_c - l^i_M = log(w^i_c) - log(w^i_M)$. The difference between log of normalized weights of conventional and modified RBPF is, $log(\hat{w}^i_{c_k}) - log(\hat{w}^i_{M_k}) = log(w^i_{c_k}) - log(\sum_{i=1}^{N}w^i_{c_k}) - log(w^i_{M_k}) + log(\sum_{i=1}^{N}w^i_{M_k})$.
There will be a maximum value among $N$ weights as $w^{m}_{c_k}$ and $w^{m}_{M_k}$ for both RBPFs. Then, $log(\sum_{i=1}^{N}w^i) = log(w^{m}) + log(1 + \sum_{i\not= m}^{N}(w^i/w^{m}))$ and the second term will have finite value as fractions will be $<1$ for finite particle size $N$. $log(\sum_{i=1}^{N}w^i_{M_k}) - log(\sum_{i=1}^{N}w^i_{c_k})$ is calculated as, $log(w^{m}_{M_k}) - log(w^{m}_{c_k}) + log(\sum_{i\not=m}^{N}w^i_{M_k}/w^m_{M_k}) -  log(\sum_{i\not=m}^{N}w^i_{c_k}w^m_{c_k}) = l^m_{c_k} - l^m_{M_k} + C_{wd}$.
From Lemma~\ref{lem:log-likelihood}, $|l^i_c - l^i_M| \le \epsilon_l$, and the difference $log(\sum_{i\not=m}^{N}w^i_{M_k}/w^m_{M_k}) -  log(\sum_{i\not=m}^{N}w^i_{c_k}/w^m_{c_k}) = C_{wd}$. Then, $log(\hat{w}^i_{c_k}) - log(\hat{w}^i_{M_k}) = l^m_{c_k} - l^m_{M_k} + l^i_{c_k} - l^i_{M_k} + C_{wd}$ and $(\hat{w}^i_{c_k}/\hat{w}^i_{M_k}) - 1 = e^{l^m_{c_k} - l^m_{M_k} + l^i_{c_k} - l^i_{M_k} + C_{wd}} - 1$ leading to,
\begin{gather*}
	|\hat{w}^i_{c_k} - \hat{w}^i_{M_k}| \le c_{w_M}(e^{2\epsilon_l + |C_{wd}|} + 1).
	\tag{16} \label{weight_update_eqn}
\end{gather*}
 where $|\hat{w}^i_{M_k}| \le c_{w_M}$ with $c_{w_M} < 1$. 
As $\hat{\theta} = \sum_{i=1}^N\theta^i\hat{w}^i$ and,
\begin{gather*}
	\|\theta^+_{c_k} - \theta^+_{M_k}\| = \|\sum_{i=1}^N\theta^i_{c_k}\hat{w}^i_{c_k} - \sum_{i=1}^N\theta^i_{M_k}\hat{w}^i_{M_k}\| \\
	\hspace*{1cm}\le \sum_{i=1}^N\|\theta^i_{c_k}\hat{w}^i_{c_k} - \theta^i_{M_k}\hat{w}^i_{M_k}\|.
\end{gather*}
Adding and subtracting $\theta^i_{c_k}\hat{w}^i_{M_k}$ in above equation, we get,
\begin{gather*}
	\sum_{i=1}^N\|\theta^i_{c_k}\hat{w}^i_{c_k} + \theta^i_{c_k}\hat{w}^i_{M_k} - \theta^i_{c_k}\hat{w}^i_{M_k} - \theta^i_{M_k}\hat{w}^i_{M_k}\| 
	= \sum_{i=1}^N\|\theta^i_{c_k}(\hat{w}^i_{c_k} \\ - \hat{w}^i_{M_k})  + (\theta^i_{c_k} - \theta^i_{M_k})\hat{w}^i_{M_k}\|. \hspace*{4cm}
\end{gather*}
$\theta^i_{c_k}$ is bounded as mentioned in Assumption$3$ and let $\epsilon_w = 2\epsilon_l + |C_{wd}|$, the posterior parameter is bounded by,
\[\|\theta^+_{c_k} - \theta^+_{M_k}\| \le \sum_{i=1}^N c_{\theta_c}(e^{\epsilon_w} + 1) + c_{w_M}\|\theta^i_{c_k} - \theta^i_{M_k}\|. \tag{17} \label{err_bound_postparam_eqn}\]
\section{BIOMOLECULAR SYSTEM EXAMPLE}
We consider gene expression model for parameter estimation using modified RBPF. A gene expression model encapsulates transcription and translation through a two-phase biochemical process. The processes are represented through CRNs to model the system dynamics conveniently. \newline
\textbf{Transcription:} The gene ($G$) binds reversibly with a free RNA polymerase ($P_o$) to form the transcription complex ($C_1$), which subsequently generates mRNA transcript ($T$):
\[
\ce{G + P_o <=>[\ce{k_{bp}}][\ce{k_{up}}] C_1}, \quad \ce{C_1 ->[\ce{k_{tx}}] G + P_o + T}.
\]
\textbf{Translation:} mRNA transcript ($T$) binds reversibly to a free ribosome ($R_i$) to produce a complex ($C_2$), which produces protein ($X$):
\[
\ce{T + R_i <=>[\ce{k_{br}}][\ce{k_{ur}}] C_2}, \quad \ce{C_2 ->[\ce{k_{tl}}] T + R_i + X}.
\]
\textbf{Degradation:} mRNA transcript and protein degrade via first-order reactions:
\[
\ce{T ->[\ce{d_T}] \varnothing}, \quad \ce{X ->[\ce{d_X}] \varnothing}.
\]
Here, $k_{bp}$, $k_{up}$, $k_{tx}$, $k_{br}$, $k_{ur}$, $k_{tl}$, $d_T$, and $d_X$ denote the corresponding kinetic parameters. Applying conservation laws for total polymerase ($P_{\text{tot}} = P_o + C_1$) and total ribosome population($R_{\text{tot}} = R_i + C_2$) along with quasi steady state assumption, the reduced order deterministic model for the system in discrete domain for sampling interval $\Delta$ is,
\begin{align*}
	T_{k+1} &= T_k + \Delta\left(k_{tx}P_{tot}\left(\frac{G}{K_1 + G} \right) - d_TT_k\right), \nonumber \\
	X_{k+1} &= X_k + \Delta\left(k_{tl}R_{\text{tot}}\left(\frac{T_k}{K_0 + T_k} \right) - d_XX_k\right), \tag{7} \label{eq:red_model}
\end{align*}
for $(k + 1)^{th}$ time instant. In~\cite{dash2025extended}, we have given a reduced order CLE model for the same gene expression system. Similarly, the stochastic dynamics represented in CLE for discrete time is given as,
\begin{gather*}
	T_{k+1} = T_k + \Delta\left(k_{tx}P_{tot}\left(\frac{G}{K_1 + G}\right) - d_TT_k\right) + \\
	\sqrt{\Delta}\left( - \sqrt{d_TT_k}\text{W}_3 +\sqrt{k_{tx}P_{tot}\left(\frac{G}{K_1 + G} \right)}\text{W}_1 \right) \\
	X_{k+1} = X_k + \Delta\left(k_{tl}R_{\text{tot}}\left(\frac{T_k}{K_0 + T_k}\right) - d_XX_k\right) + \\
	\sqrt{\Delta}\left(- \sqrt{d_XX_k}\text{W}_4 +   \sqrt{k_{tl}R_{\text{tot}}\left(\frac{T_k}{K_0 + T_k} \right)}\text{W}_2 \right),  \tag{8}
	\label{eq:cle_eqn}
\end{gather*}
where $\text{W}_i \sim \mathcal{N}(0, 1)$ are independent Gaussian noise terms, $K_0 = \frac{k_{tl} + k_{ur}}{k_{br}}$ and $K_1 = \frac{k_{tx} + k_{up}}{k_{bp}}$.
\section{RESULTS}
We first find suitable identifiable parameters which can be estimated through sensitivity analysis, and then generate system data using stochastic simulation algorithm (SSA). The system parameters taken for sensitivity analysis and SSA are $P_{tot} = 100$, $R_{tot} = 100$, $G = 10$, $k_{tx} = k_{tl} = 0.1$, $k_{up} = k_{ur} = 0.05$, $k_{bp} = k_{br} = 0.02$, $d_T = 0.01$, $d_X = 0.01$. After data generation we implement modified RBPF with mRNA transcript $T$ and protein $X$ as measurements and give comparison between conventional RBPF for parameter estimation, autocorrelation test and normalized root mean square error (NRMSE).
\subsection{Sensitivity Analysis}
\begin{figure}[thpb]
	\begin{subfigure}{42mm}
		\includegraphics[scale=0.26]{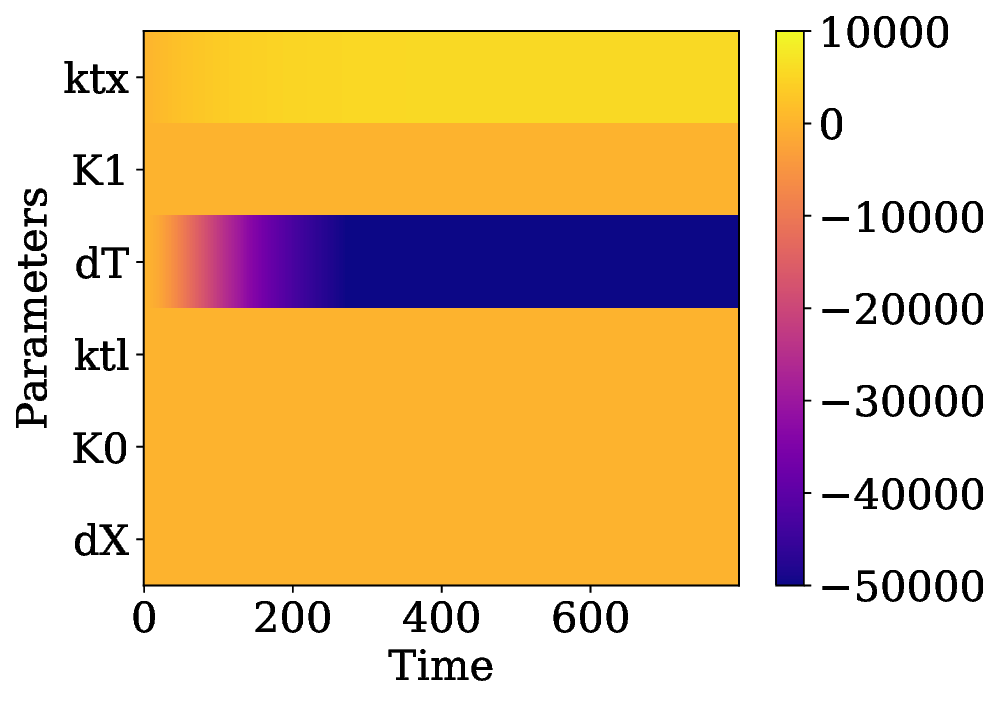}
		\caption{}
	\end{subfigure}
	\begin{subfigure}{42mm}
		\includegraphics[scale=0.26]{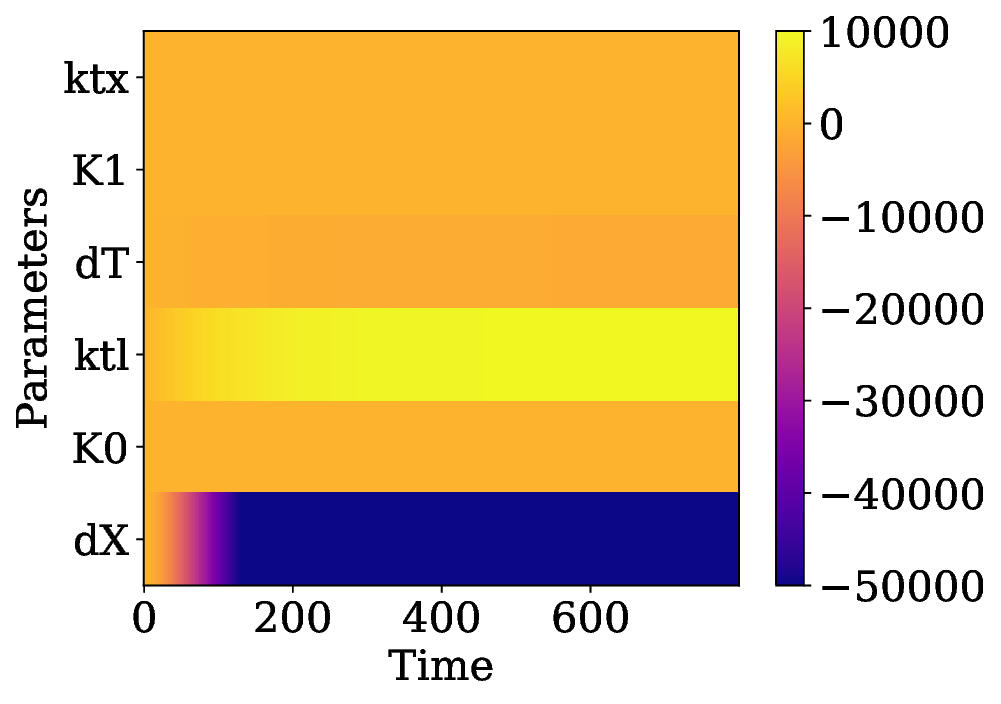}
		\caption{}
	\end{subfigure}
	\caption{Sensitivity analysis for (a) $T$ and (b) $X$ state}
	\label{sensitivity_analysis}
\end{figure}

To determine which parameters can be identified and estimated, sensitivity analysis of parameter with respect to measurements is taken. Using ~\textit{biocrnpyler}~\cite{poole2022biocrnpyler} and \textit{bioscrape}~\cite{Swaminathan2023} package in python for reduced order gene expression model, we perform the sensitivity analysis and the results are shown in Fig.~\ref{sensitivity_analysis} and tabulated in Table.~\ref{tab:sensitivity_analysis}. Degradation parameters $d_X$ and $d_T$ show highest sensitivity which is valid as mathematically these parameters are directly related to change in state, followed by $k_{tl}$ and $k_{tx}$. So, using measurements which denote the states itself in this case, $d_X$ and $d_T$ were calculated using graphical method~\cite{alon2019introduction},
\[
d_X = \frac{log(2)}{t(X)_{\frac{1}{2}}}, \hspace*{1cm}
d_T = \frac{log(2)}{t(T)_{\frac{1}{2}}}. \tag{9} \label{dis_eqn}
\]
$t(.)_{\frac{1}{2}}$ denotes the time taken to reach halfway between the initial and final concentration. Mathematically we take, time taken to reach double of initial concentration.
\begin{table}[thpb]
	\caption{SENSITIVITY ANALYSIS OF MEASURED STATES WITH RESPECT TO PARAMETERS}
	\begin{tabular}{|m{0.85cm}|m{0.8cm}|m{0.8cm}|m{0.8cm}|m{0.7cm}|m{1cm}|m{0.85cm}|}
		\hline
		\multirow{2}{1cm}{Outputs} & \multicolumn{6}{c|}{Sensitivity} \\
		\cline{2-7}
		& $k_{tx}$ & $K_1$ & $k_{tl}$ & $K_0$ & $d_T$ & $d_X$ \\
		\hline
		T & 5712.37  & -32.64 & 0 & 0 & -57104 & 0\\
		\hline
		X & 127.57 & -0.73 & 9867.1 & -1.71 & -1275.3 & -98633\\
		\hline
	\end{tabular}
	\label{tab:sensitivity_analysis}
\end{table}

\subsection{Data Generation}
The datasets for estimation are generated through SSA which takes care of intrinsic noise in CRNs, the process noise in Kalman filter. $50$ datasets were generated and measurements contained the observed state along with measurement noise using \textit{randint} in MATLAB of range $[-1,1]$ for $T$ state and $[-3,3]$ for $X$ state. 

\begin{figure}[thpb]
	\begin{subfigure}{42mm}
		\includegraphics[scale=0.3]{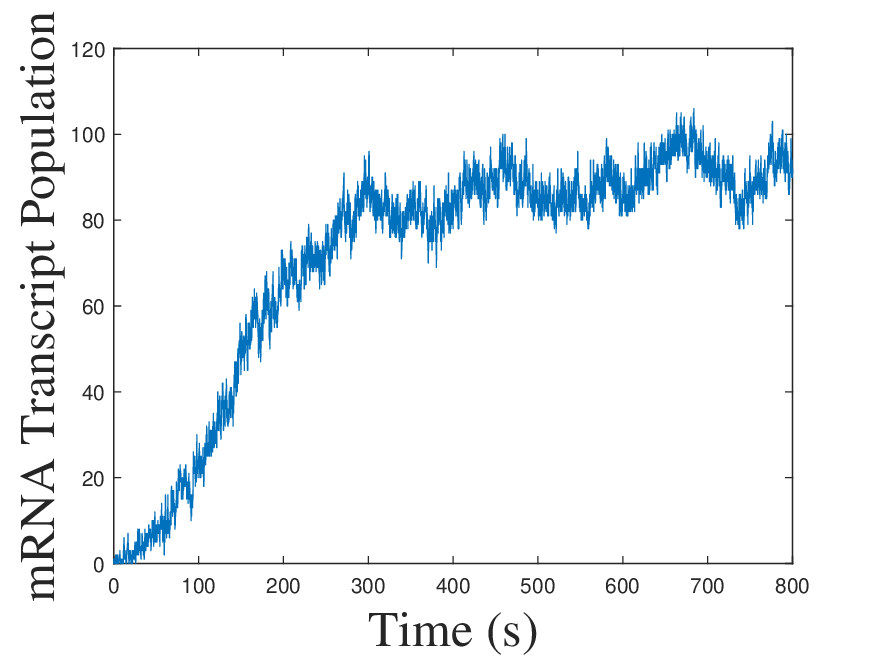}
		\caption{}
	\end{subfigure}
	\begin{subfigure}{42mm}
		\includegraphics[scale=0.3]{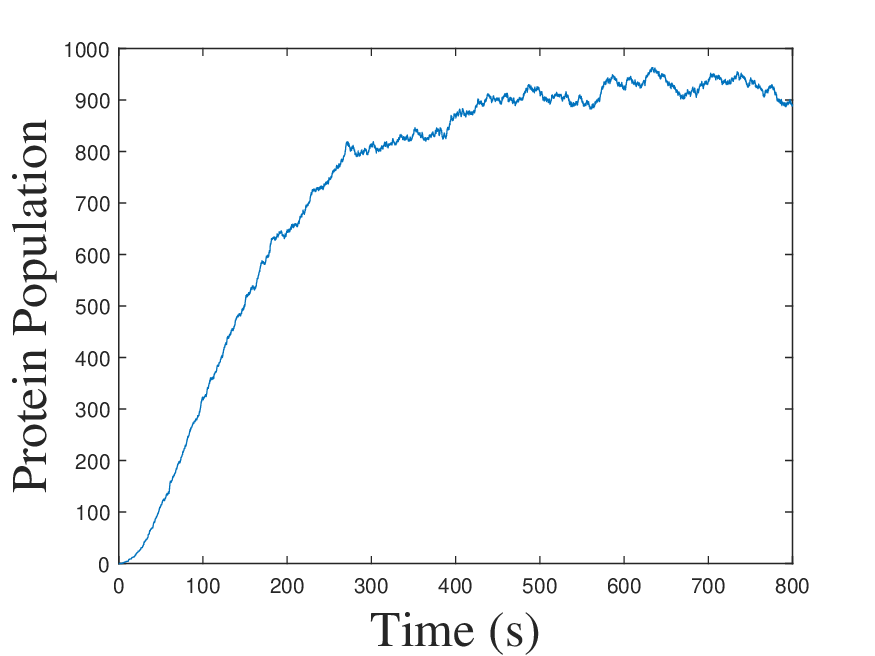}
		\caption{}
	\end{subfigure}
	\caption{A (a) mRNA transcript (T) and (b) protein (X) dataset generated for the gene expression system through SSA}
	\label{ssa}
\end{figure}

\subsection{Modified RBPF} 
\begin{figure}[thpb]
	\centering
	\begin{subfigure}{42mm}
		\includegraphics[scale=0.3]{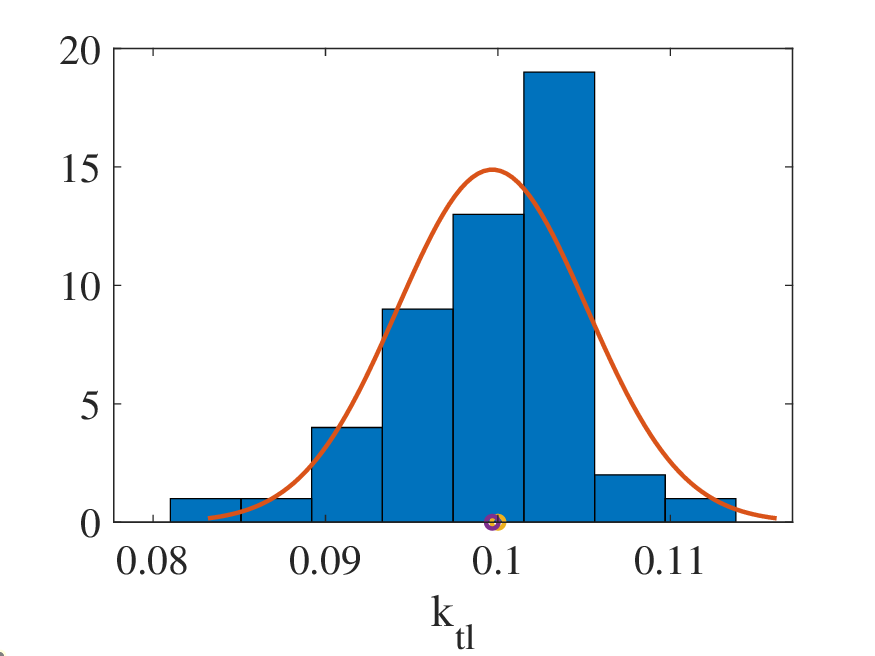}
		\caption{}
	\end{subfigure}
	\begin{subfigure}{42mm}
		\includegraphics[scale=0.3]{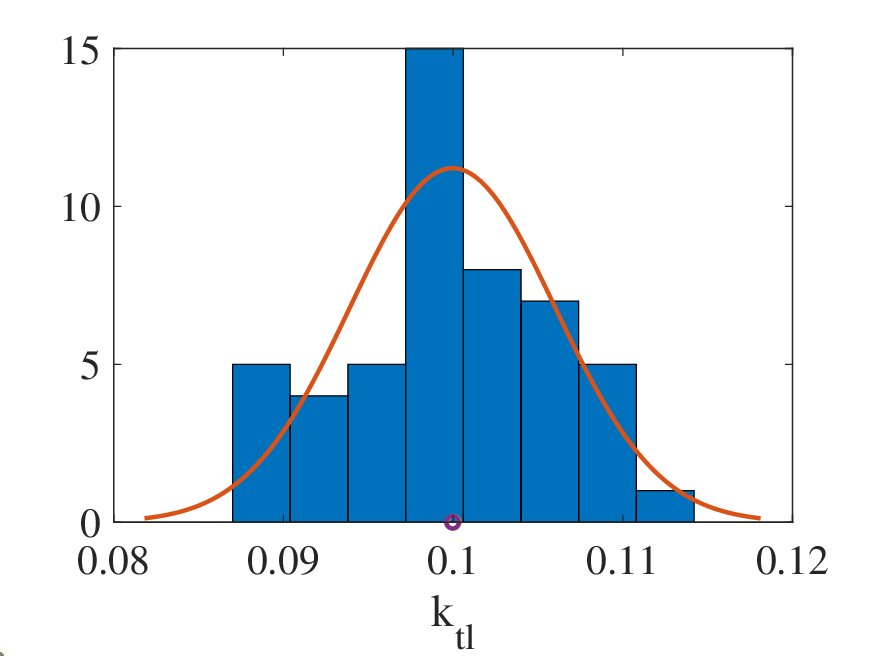}
		\caption{}
	\end{subfigure}
	\begin{subfigure}{42mm}
		\includegraphics[scale=0.3]{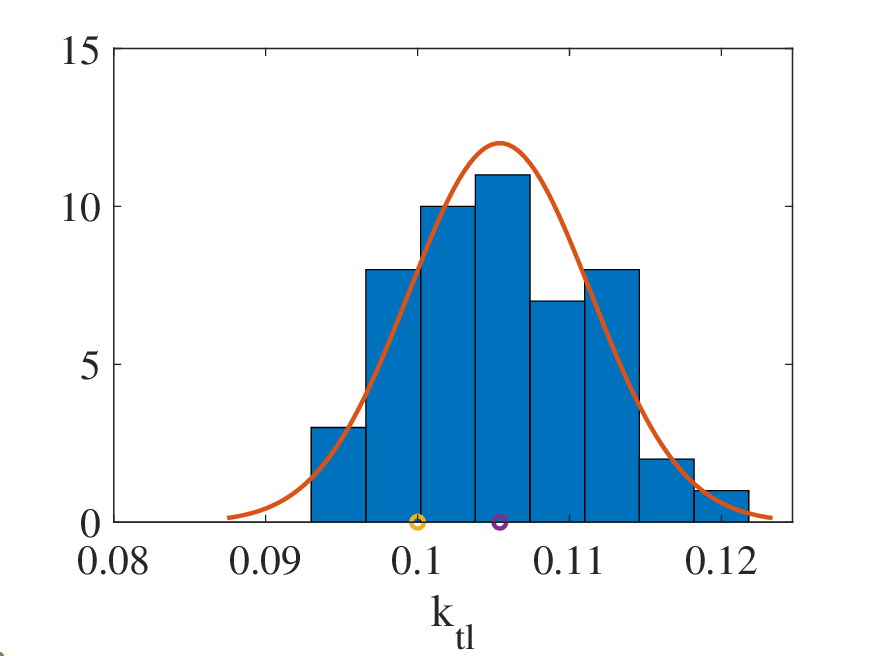}
		\caption{}
	\end{subfigure}
	\begin{subfigure}{42mm}
		\includegraphics[scale=0.3]{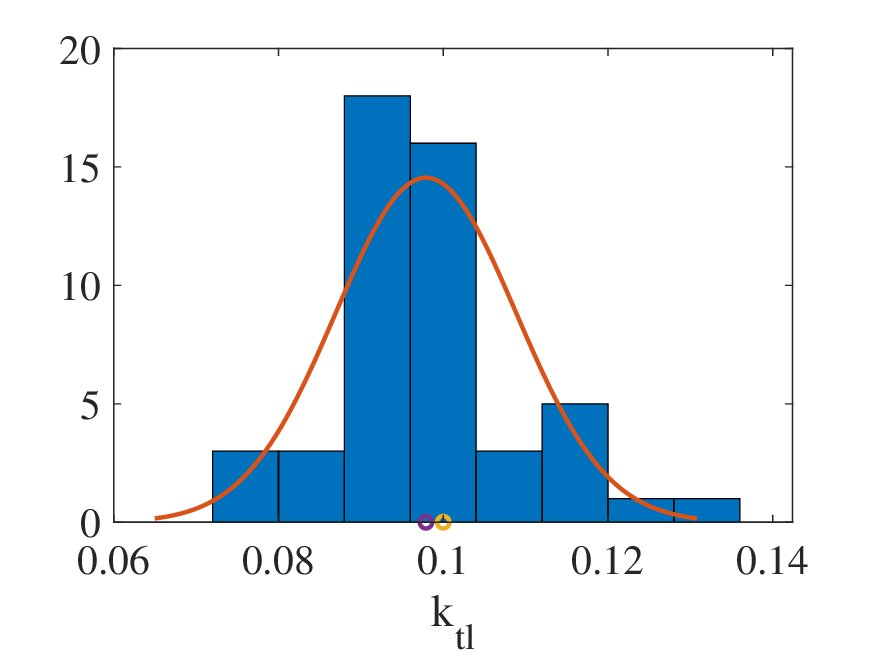}
		\caption{}
	\end{subfigure}
	\begin{subfigure}{42mm}
		\includegraphics[scale=0.3]{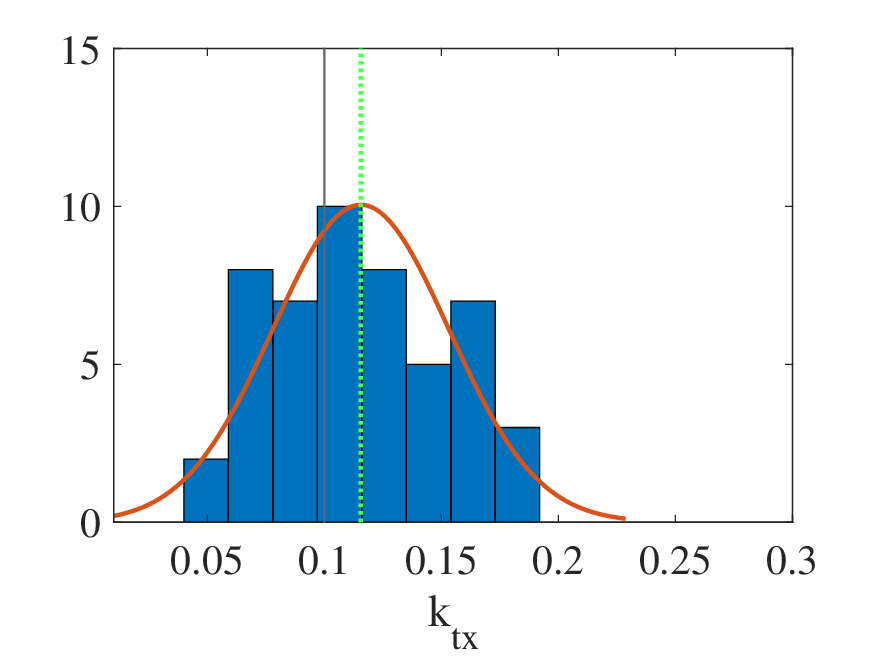}
		\caption{}
	\end{subfigure}
	\begin{subfigure}{42mm}
		\includegraphics[scale=0.3]{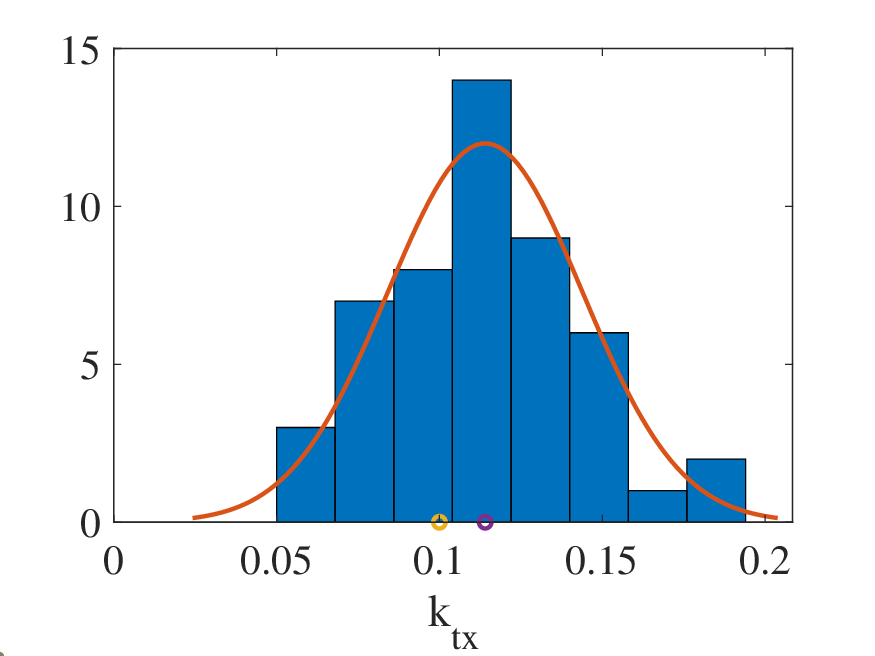}
		\caption{}
	\end{subfigure}
	\begin{subfigure}{42mm}
		\includegraphics[scale=0.3]{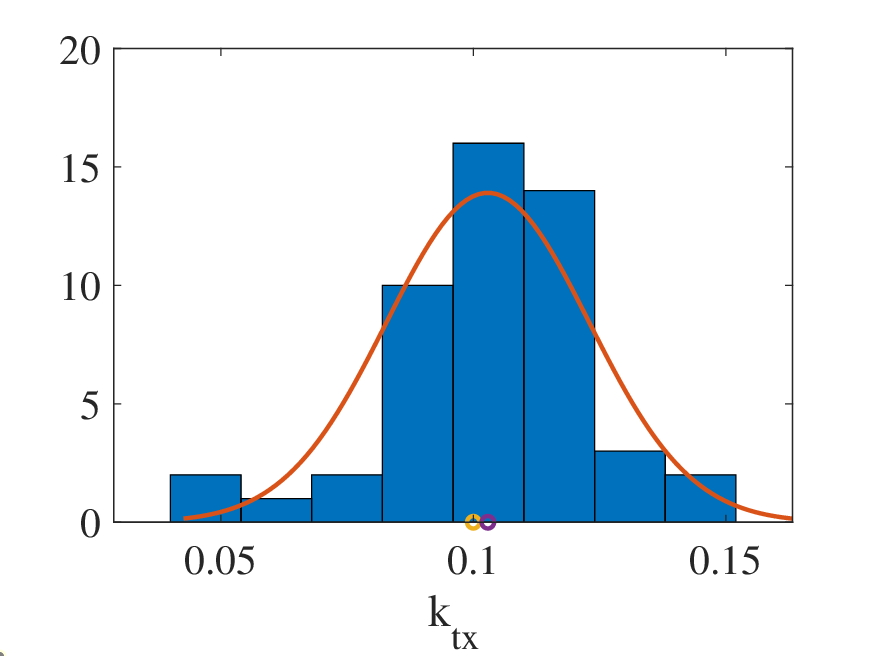}
		\caption{}
	\end{subfigure}
	\begin{subfigure}{42mm}
		\includegraphics[scale=0.3]{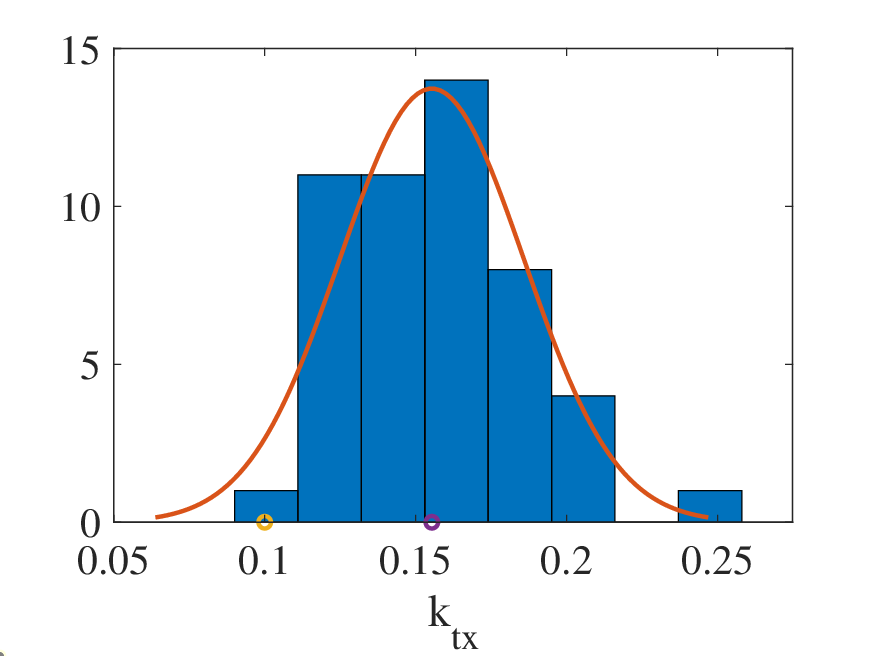}
		\caption{}
	\end{subfigure}
	\caption{Histogram plots of estimated $k_{tl}$ for (a) $Q = 0.1 \times I_{n \times n}$, (b) $Q=1 \times I_{n \times n}$, (c) $\hat{Q}_k$, (d) $Q=10 \times I_{n \times n}$ and estimated $k_{tx}$ for (e) $Q = 0.1 \times I_{n \times n}$, (f) $Q=1 \times I_{n \times n}$, (g) $\hat{Q}_k$, (h) $Q=10 \times I_{n \times n}$ using modified RBPF where solid line denotes true $k_{tl}=0.1$ and $k_{tx}=0.1$. Dotted line denotes estimated $k_{tl}$ mean and $k_{tx}$ mean across 50 datasets.}
	\label{ktl_ktx_histogram}
\end{figure}

\begin{figure}[thpb]
	\centering
	\begin{subfigure}{40mm}
		\includegraphics[scale=0.29]{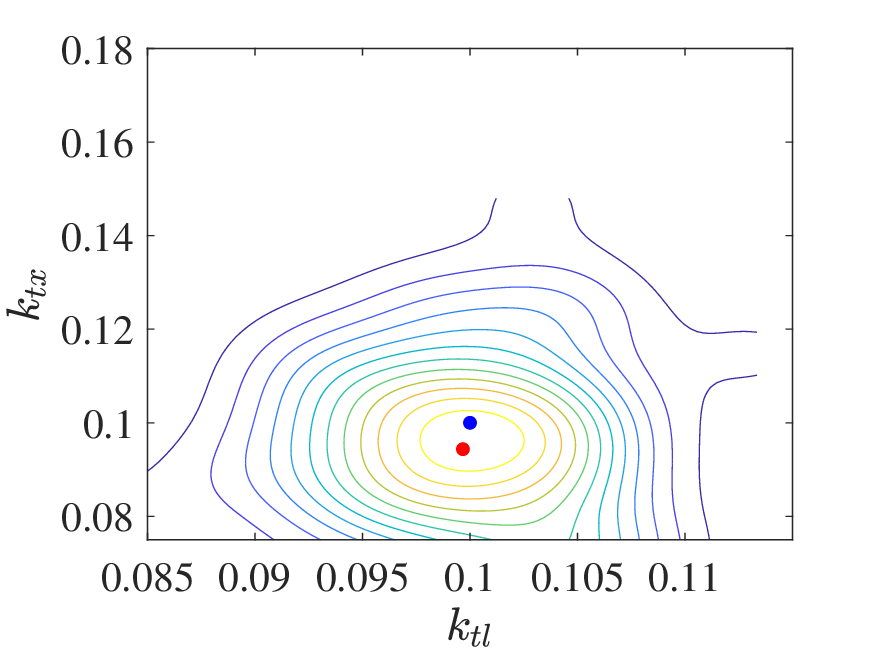}
		\caption{}
	\end{subfigure}
	\begin{subfigure}{40mm}
		\includegraphics[scale=0.29]{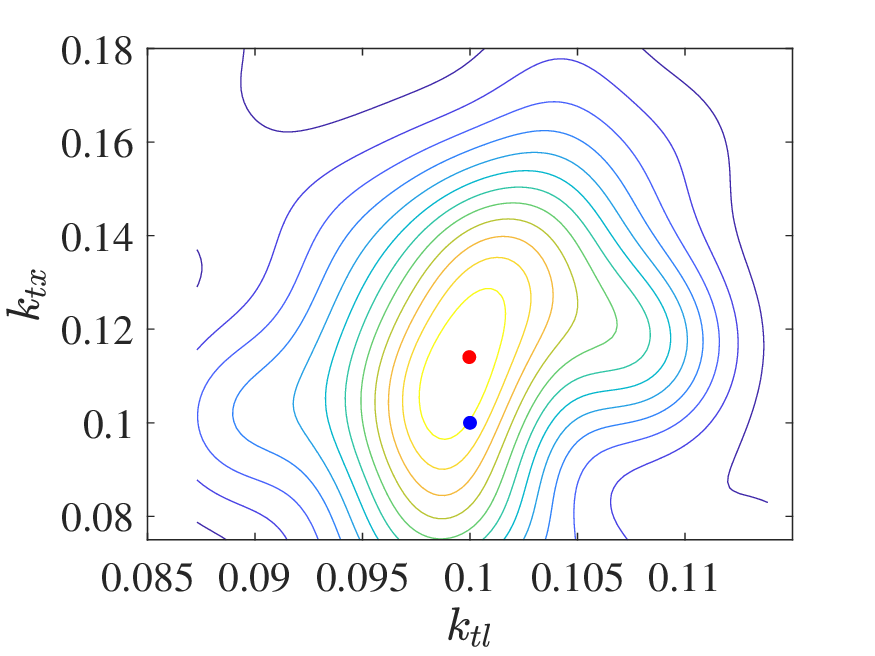}
		\caption{}
	\end{subfigure}
	\begin{subfigure}{40mm}
		\includegraphics[scale=0.29]{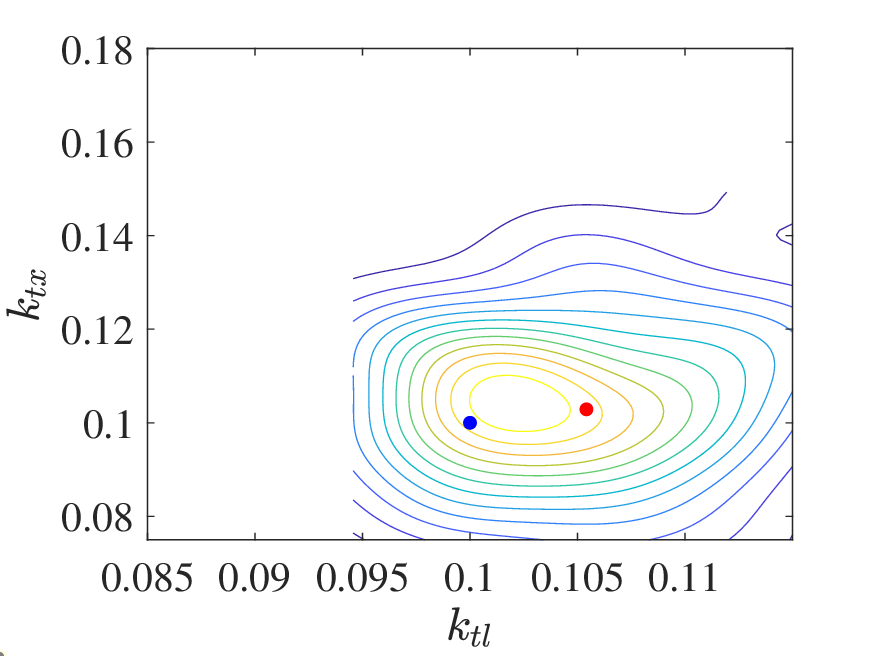}
		\caption{}
	\end{subfigure}
	\begin{subfigure}{40mm}
		\includegraphics[scale=0.29]{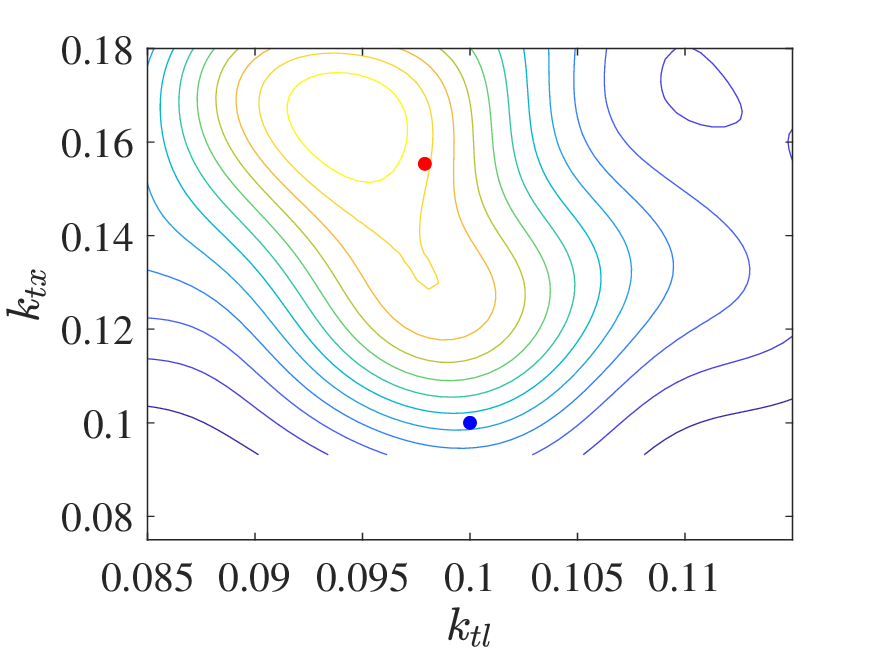}
		\caption{}
	\end{subfigure}
	\caption{Contour plot of estimated $k_{tl}$ and $k_{tx}$ bivariate distribution for (a)$Q = 0.1 \times I_{n \times n}$, (b) $Q=1 \times I_{n \times n}$, (c) $\hat{Q}_k$, (d) $Q=10 \times I_{n \times n}$ using modified RBPF where blue marker is the estimated mean parameter and red marker is the true parameter. Distance between markers are (a) 0.006 (b) 0.014 (c) 0.006 (d) 0.055 respectively. Plots are generated using a grid of parameter mean across 50 datasets and apply \textit{mvksdensity} function in MATLAB to obtain probability density function to represent contour lines.}
	\label{contour_plot_2D_dist}
\end{figure}

Given measurements $T$ and $X$, we calculate $d_T$ and $d_X$ using graphical methods given in~\eqref{dis_eqn}. After finding estimated $d_T = 0.0101$ and $d_X = 0.0099$ we use modified RBPF in discrete domain with $\Delta = 0.08$ to estimate states and parameters $k_{tl}$ and $k_{tx}$ for $50$ datasets of $800$ seconds duration. State vector is $\zeta_k = [T_k, X_k]^T$ and parameter vector is $\theta_{k} = [k_{tl}, k_{tx}]^T$. Initial state is $\zeta_0 = [0, 0]^T$ and parameters are initially picked from uniform distribution where range is between half and double of nominal parameter i.e, $k_{tl} \sim [0.05, 0.2]$, $k_{tx} \sim [0.05, 0.2]$. We have $C = I_{2 \times 2}$ with measurement noise covariance $R = diag(1,5)$ and initial error covariance matrix $P^+_{0} = 5I_{n \times n}$. The adaptive resampling threshold $N^0_{eff} = 0.98N$ where $N = 100$ represents number of particles and covariance threshold $\Sigma_{lb} = diag(10^{-7}, 10^{-6})$.
The estimation performance is compared with constant process noise covariances, $Q=0.1 \times I_{n \times n}$, $Q=1 \times I_{n \times n}$, $Q=10 \times I_{n \times n}$ to estimate-dependent $\hat{Q}_k$ between modified RBPF, conventional RBPF and CD-EKF with updating process noise covariance represented in notations $(.)(M)$, $(.)(c)$ and $(.)(E)$ respectively. As CD-EKF augments parameter vector as states to estimate, it has same specification as that of compared RBPFs, but the initial parameter vector is defined as $[0.125, 0.125]^T$. Histogram plots show the distribution of estimated parameters using modified RBPF across 50 datasets for different process noise covariance in Fig.~\ref{ktl_ktx_histogram}.
 The mean and standard deviation of estimated parameters across all datasets are tabulated in Table.~\ref{tab:est_param}. Estimated values in case of modified RBPF are more accurate as compared to conventional RBPF for all process noise covariance selections except $k_{tx}$ mean in $Q=10I_{n \times n}$. For estimate-dependent $Q$ ($\hat{Q}_k$) both parameter estimates are close to estimates obtained in case $Q =0.1I_{n \times n}$. While CD-EKF gives accurate estimation of $k_{tl}$ for all process noise covariance, but $k_{tx}$ estimation is less accurate as compared to modified RBPF.
\begin{table}[thpb]
	\caption{PARAMETER ESTIMATION WITH DIFFERENT PROCESS NOISE COVARIANCE FOR MODIFIED AND CONVENTIONAL RBPF}
	\begin{tabular}{|p{0.75cm}|p{0.5cm}|p{0.5cm}|p{0.5cm}|p{0.5cm}|p{0.5cm}|p{0.5cm}|p{0.5cm}|p{0.5cm}|}
		\hline
		\multirow{2}{0.75cm}{Para-meter} &  \multicolumn{2}{m{1cm}|}{$Q = 0.1\times I_{n\times n}$} & \multicolumn{2}{m{1cm}|}{$Q = I_{n\times n}$ } & \multicolumn{2}{m{1cm}|}{$\hat{Q}_k$} & \multicolumn{2}{m{1cm}|}{$Q = 10\times I_{n\times n}$} \\
		\cline{2-9}
		& Mean & Std Dev & Mean & Std Dev & Mean & Std Dev & Mean & Std Dev \\
		\hline
		$k_{tl}(M)$ & 0.099 & 0.005 & 0.1 & 0.006 & 0.105 & 0.006 & 0.098 & 0.011 \\
		\hline
		$k_{tl}(c)$ & 0.097 & 0.006 & 0.101 & 0.009 & 0.113 & 0.035 & 0.119 & 0.032\\
		\hline
		$k_{tl}(E)$ & 0.1 & 0.005 & 0.1 & 0.005 & 0.099 & 0.003 & 0.099 & 0.003 \\
		\hline
		$k_{tx}(M)$ & 0.094 & 0.018 & 0.114 & 0.029 & 0.103 & 0.019 & 0.155 & 0.03\\
		\hline
		$k_{tx}(c)$& 0.053 & 0.022 & 0.122 & 0.027 & 0.17 & 0.056 & 0.14 & 0.043\\
		\hline
		$k_{tx}(E)$ & 0.079 & 0.012 &0.079 & 0.012 & 0.084 & 0.007 & 0.076 & 0.004\\
		\hline
	\end{tabular}
	\label{tab:est_param}
\end{table}
Contour plots of estimated $k_{tl}$ and $k_{tx}$ as a bivariate distribution using modified RBPF is also plotted for different process noise covariance in Fig.~\ref{contour_plot_2D_dist}. The true value is nearer to the mean of distribution in case of $\hat{Q}_k$ and $Q=0.1I_{n \times n}$ as compared to other fixed Q values.

NRMSE between measurements of size $N_T$ and corresponding estimated outputs is calculated as, $$NRMSE(.) = \frac{\sqrt{\frac{\sum_{i=1}^{N_T}(y_i - C\zeta_i)^2}{N_T}}}{\frac{\sum_{i=1}^{N_T}y_i}{N_T}}.$$ The NRMSE results for both modified RBPF and conventional RBPF shown in Table.~\ref{tab:nrmse}. $NRMSE(T)$ and $NRMSE(X)$ for modified RBPF is more compared to conventional RBPF. As shown in Section~\MakeUppercase{\romannumeral 3}, since state has inter-algorithm deviation, the NRMSE is also affected where we see some finite difference of performance between two algorithms. As compared CD-EKF, NRMSE is comparable or slightly less.

\begin{table}
	\caption{NRMSE COMPARISON BETWEEN MODIFIED AND CONVENTIONAL RBPF OF $T$ AND $X$ FOR DIFFERENT PROCESS NOISE COVARIANCE}
	\begin{tabular}{|p{0.6cm}|p{0.5cm}|p{0.5cm}|p{0.5cm}|p{0.5cm}|p{0.5cm}|p{0.5cm}|p{0.5cm}|p{0.5cm}|}
		\hline
		\multirow{2}{0.6cm}{State} &  \multicolumn{2}{p{1cm}|}{$Q=0.1\times I_{n\times n}$} & \multicolumn{2}{p{1cm}|}{$Q=I_{n\times n}$ } & \multicolumn{2}{p{1cm}|}{$\hat{Q}_k$} & \multicolumn{2}{p{1cm}|}{$Q=10\times I_{n\times n}$} \\
		\cline{2-9}
		& Mean & Std dev & Mean & Std dev & Mean & Std dev & Mean & Std dev \\
		\hline
		$T(M)$ & 0.02 & 0.001 & 0.009 & $4.6\times 10^{-4}$ & 0.018 & 0.001 & 0.002 & $9.9\times 10^{-4}$ \\
		\hline
		$T(c)$& $3.8 \times 10^{-4}$ & 0.003 &$1.7 \times 10^{-4}$ & 0.001 &$3.3 \times 10^{-4}$ & 0.002 & $3.8 \times 10^{-4}$ & $2.7 \times 10^{-4}$\\
		\hline
		$T(E)$ & 0.027 & 0.001 & 0.027 & 0.001 & 0.02 & 0.016 & 0.012 & $5.8 \times 10^{-4}$\\
		\hline 
		$X(M)$ & 0.004 & $6.8\times 10^{-5}$ & 0.002 & $4.8\times 10^{-5}$ & 0.002 & $5.8\times 10^{-5}$ & 0.001 & $2.2\times 10^{-5}$\\
		\hline
		$X(c)$ & $6.9 \times 10^{-4}$ & $4.9 \times 10^{-4}$ & $4.5 \times 10^{-4}$ & $3.1 \times 10^{-4}$ & $4.2 \times 10^{-4}$ & $2.9 \times 10^{-4}$ & $1.9 \times 10^{-4}$ & $1.3 \times 10^{-4}$ \\
		\hline
		$X(E)$ & 0.004 & $10^{-4}$ & 0.004 & $10^{-4}$ & 0.004 & 0.007 & 0.002 & $4\times 10^{-5}$ \\
		\hline
	\end{tabular}
	\label{tab:nrmse}
\end{table}

To check the optimality of filters we performed whiteness test (\cite{mehra1970identification}) for both observed states $T$ and $X$ with $95\%$ confidence limits with results shown in Table.~\ref{tab:whiteness_test}. Number of lags outside the confidence limits for both $T$ and $X$ is less in modified RBPF as compared to conventional RBPF. In case of $\hat{Q}_k$, number of lags in $T$ is not good as compared to $Q = I_{n \times n}$ and $Q = 10I_{n \times n}$, whereas number of lags is minimum for $X$ state. Whiteness tests of CD-EKF show that modified RBPF have fewer lags for both $T$ and $X$ states. 
\begin{table}[thpb]
	\caption{WHITENESS TEST OF OBSERVED STATES FOR DIFFERENT PROCESS NOISE COVARIANCE FOR 50 DATASETS BETWEEN MODIFIED AND CONVENTIONAL RBPF}
	\begin{tabular}{|m{2cm}|m{1cm}|m{1cm}|m{1cm}|m{1cm}|}
		\hline
		Number of correlation lags outside confidence limits & $Q = 0.1\times I_{n\times n}$ & $Q = I_{n\times n}$ & $\hat{Q}_k$ & $Q = 10\times I_{n\times n}$\\
		\hline
		$T(M)$-proposed & 19 & 9 & 19 & 5 \\
		\hline
		$T(c)$ & 19 & 9 & 88 & 5 \\
		\hline
		$T(E)$ & 99 & 99 & 82 & 40 \\
		\hline
		$X(M)$-proposed & 23 & 4 & 2 & 2 \\
		\hline
		$X(c)$ & 100 & 5 & 100 & 99 \\
		\hline
		$X(E)$ & 100 & 100 & 36 & 12\\
		\hline		
	\end{tabular}
	\label{tab:whiteness_test}
\end{table} 

\section{Conclusion}
A modified RBPF was implemented for the reduced order gene expression system, where possibility of multiple unknown parameters exist. Sensitivity analysis helped in obtaining identifiable parameters and were estimated through modified RBPF with EKF based on updating process noise covariance matrix. We compared the accuracy of parameter estimates and optimality of filter performance for the proposed modified RBPF with a conventional and EKF baseline with fixed choices of process noise covariance. We observed that the accuracy of parameter estimates for updating noise covariance RBPF is comparable to the best performing cases of fixed noise covariance. We know that increasing the process noise covariance reduces the state estimation NRMSE as the EKF part relies more on measurements than the model. This trend is also reflected in the innovation whiteness test where the number of autocorrelation coefficients outside $95\%$ confidence limit reduces substantially for $Q = I_{n \times n}$ and $10I_{n \times n}$ for the transcript ($T$) measurement. For the protein measurement ($X$), RBPF with updated Q produces similar whiteness results as in $I_{n \times n}$ and $10I_{n \times n}$, significantly outperforming $0.1I_{n \times n}$. However, these improved optimality results do not necessarily imply better parameter estimates especially for the $k_{tx}$ parameter which is primarily inferred from the $T$ measurements. A larger process noise covariance reduces the estimation error and produces more white innovation sequence through tighter state correction in EKF, and thereby reduces the ability of particle filter to distinguish between likelihood of competing parameter particles. In contrast, the RBPF with updated process noise covariance based on first principles of CLE achieves parameter estimation accuracy comparable to the best cases of fixed noise covariance and simultaneously maintains satisfactory innovation whiteness characteristics. Also comparing with conventional RBPF and CD-EKF, parameter estimates and whiteness tests for updating process noise covariance are better in case of modified RBPF. NRMSE is lower in conventional RBPF, but the proposed RBPF still has acceptable NRMSE values. These results indicate that the proposed filter provides a balance between optimality of the state EKF and accuracy of the parameter PF in the RBPF framework along with reduced computational burden as compared to conventional RBPF.



\bibliography{refs}             
                                                   







\end{document}